\documentclass[9pt,twocolumn,twoside]{osajnl}
\definecolor{mypink3}{cmyk}{0, 0.7808, 0.4429, 0.1412}
\usepackage{caption}
\usepackage{subcaption}
\usepackage[font=small,labelfont=bf,
   justification=justified,
   format=plain]{caption}
   
\usepackage{xcolor}
\newcommand \red[1]{#1}
   
\journal{ol}
\setboolean{shortarticle}{true}

\title{Impact of FBG feedback phase on laser dynamics}

\author[1]{Martin Skënderas}
\author[1]{Spencer W. Jolly}
\author[1]{Nitish Gupta}
\author[1]{Thomas Geernaert}
\author[1]{Martin Virte}

\affil[1]{Brussels Photonics (B-PHOT), Vrije Universiteit Brussel and Flanders Make, Dept. Of Applied Physics and Photonics, Pleinlaan 2, B-1050 Brussels, Belgium}

\begin{abstract}
Fiber Bragg Gratings have been advantageously used to improve the chaotic properties of semiconductor lasers. Though these components are known to be highly sensitive to environmental conditions, feedback phase fluctuations are often neglected. In this work, we experimentally demonstrate that the small variations of the propagation time induced by a simple thermal tuning of the FBG is sufficient to induce significant changes of the laser behavior. We report periodic stability enhancements linked with phase variations and highlight that both phase variation and phase offsets play an important role. Last, we show a good qualitative agreement with simulations based on an expanded version of the Lang-Kobayashi model. 
\end{abstract}

\setboolean{displaycopyright}{true}

\begin{document}

\maketitle

The dynamical behavior and the properties of a semiconductor laser can be notably affected by an external optical feedback and will be highly sensitive to all kinds of feedback features. Optical feedback typically compromises the performance of a laser but is also the source of several applications which have in their foundation the variety of the complex dynamics exhibited by the laser. Amid these, secure communication \cite{argyris_chaos-based_2005}, fast physical random-number generation \cite{uchida_fast_2008}, remote chaotic lidar/ radar \cite{1303798,Tsay:20} and optical time-domain reflectometry \cite{4581643}.

Recently, Fiber Bragg Gratings (FBGs) have been investigated as promising candidates to enhance the performance of such applications\red{\cite{Davis:04,Baladi:16}}. Thus far, the implementation of FBG feedback has been focused on the reduction of the time-delay signature (TDS) as shown in \cite{7097638}. This is attributed to the distribution in time of the reflections of the FBG, which are associated with the chromatic dispersion. Chirping the FBG period \cite{Wang:17,Wang:19} can make the feedback delay a key parameter to optimise and better suppress the TDS compared with the conventional FBGs. 
Recently, it has been demonstrated that TDS suppression and bandwidth enhancement are achieved for nanolasers via FBG feedback \cite{Jiang:21}. On other studies, changing the setup configuration such as introducing FBG feedback under optical injection \cite{Zhang:18} and mutual injection \cite{Wang:199} can modify the feedback features to benefit many applications via the enhancement of the chaotic dynamics. 

The emission dynamics also depend heavily on the feedback phase, that is, the phase that the light accumulates while traveling through the feedback loop. It has been shown that in the short cavity regime the dynamics evolving under the variation of the feedback phase exhibit a cyclic behavior \cite{PhysRevE.67.066214}. The effect of feedback phase on the relaxation oscillations has been studied as well, in \cite{Liu:211} it is demonstrated that the relaxation oscillation frequency (ROF) changes in a sinusoidal manner with respect to the feedback phase. The impact of phase for the case of filtered optical feedback (FOF), where the filter alters the spectral content of the feedback light, has also been investigated. FOF lasers show different types of dynamics and dependence on the feedback phase. In \cite{PhysRevE.76.026212} several regimes were identified as function of the feedback rate. FBG feedback is a specific case of FOF, however, the impact of FBG feedback phase has not been thoroughly investigated yet. 

In this letter the dependence of a semiconductor laser operation on the phase of time-distributed feedback from an FBG is studied. Experimental maps based on highly resolved optical spectra of the laser output are used to obtain a detailed overview of the laser dynamics. The maps are established in the parameter space of wavelength detuning and feedback strength in a fully automated setup. Similar maps are generated numerically using a modified version of the Lang-Kobayashi equations and we obtain an excellent qualitative agreement with the experimental data. It is found that when the Bragg wavelength of the FBG is varied a phase change is induced which modifies the laser's emission dynamics.

The FBG is fabricated in the core region of a conventional single-mode fiber (Corning SMF-28) with a standard phase mask technique using a $193$\,nm ArF excimer pulsed laser. \red{The grating has a maximum reflectivity of $90$\% and length of $4$\,mm.} The configuration of the experimental setup is shown schematically in Fig. 1(a). We use a single-mode distributed feedback laser (Thorlabs L1550P5 DFB) which is coupled into a single-mode fiber connected to an FBG, that provides optical feedback into the laser. The lasers temperature is fixed at $T=22^{\circ}$\,C by a TEC controller. To ensure that the reflected light returns towards the laser with the same polarization orientation, a free-space linear polarizer is combined with a fiber-based polarization controller. The feedback strength is controlled via a fiber-based variable optical attenuator (VOA). Single-mode fibers with angle-polished tips and narrow key mating sleeve connectors with index matching gel are used in the feedback arm to avoid unwanted reflections. \red{The overall cavity length is $5.52$\,m ($0.42$\,m free space and $5.1$\,m fiber setup) and the estimated coherence length of the laser is $2.83$\,m.} The beam splitter diverts $25$\% of the emitted light towards the measurement arm composed of a free-space optical isolator followed by a high-resolution OSA (APEX 2083A) that uses a resolution of $5$\,MHz to record the optical spectra. \red{Due to the long feedback cavity and despite the high-resolution, external cavity modes cannot be resolved.} 
\begin{figure}[htbp]
     \begin{subfigure}[b]{\linewidth}
        \includegraphics[width=\linewidth]{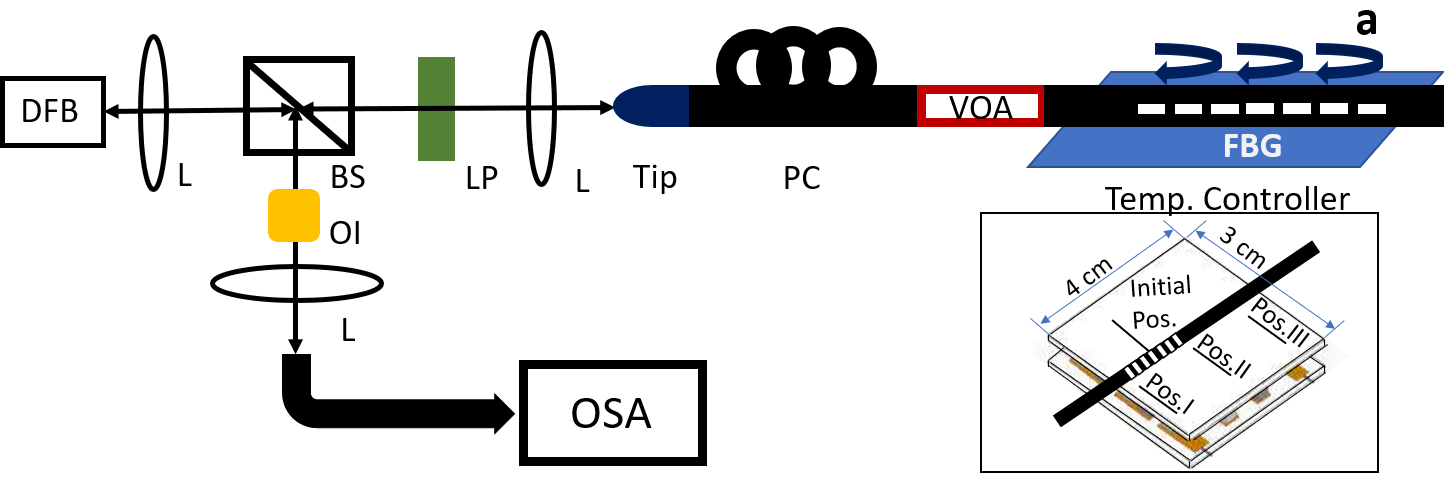}
     \end{subfigure}
     \hfill
     \begin{subfigure}[b]{\linewidth}
         \includegraphics[width=\linewidth]{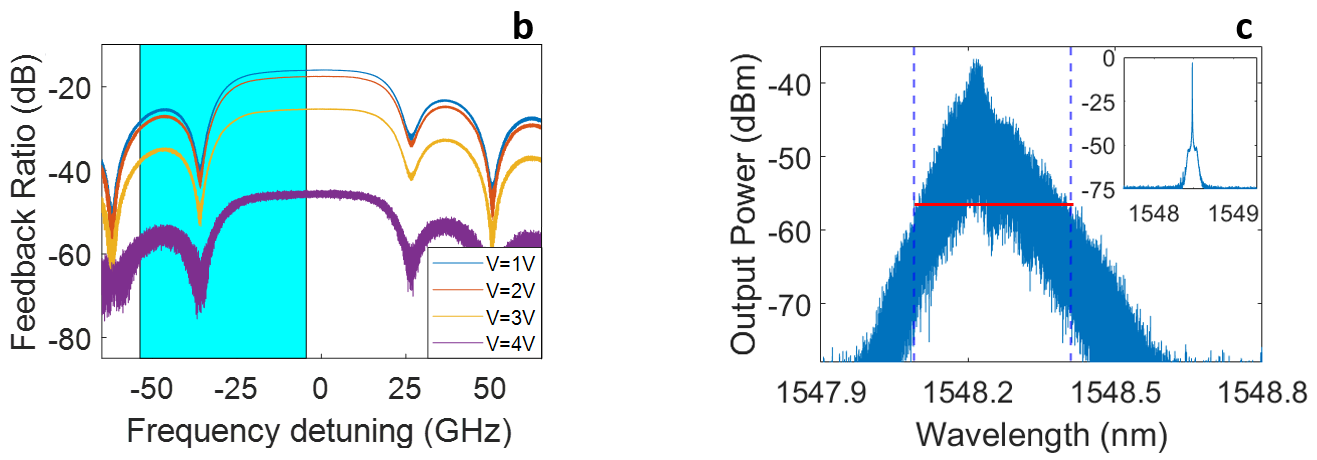}
     \end{subfigure}
        \caption{(a) Schematics of the setup. L: lens, LP: linear polarizer, BS: beam splitter, PC: polarization controller, VOA: variable optical attenuator, OI: optical isolator, OSA: optical spectrum analyser, Thick lines: optical fibers, Thin lines: free-space optical paths. \red{Inset shows the position of the grating on the heater.} \red{(b) Estimated feedback ratio for four different VOA voltages. The highlighted section shows the detuning range achieved through thermal control of the FBG.} (c) Illustration of the optical spectra analysis to extract the bandwidth at -20\,\red{dB}. In inset is the laser emission optical spectrum in stable state with injection current $26$\,mA.}
        \label{fig:setup_schematics}
\end{figure}

\red{The Bragg wavelength of the FBG is temperature-controlled in a closed loop via a Peltier element (Thorlabs TECD14) combined with a $10$\,k$\Omega$ thermistor. The grating is positioned at different positions on the heater as illustrated in Fig. 1(a)}. In our experiment, the temperature of the Peltier is increased in steps of of $0.2^{\circ}$\,C from $10^{\circ}$\,C to $45^{\circ}$\,C to detune the Bragg wavelength in a range of $48.5$\,GHz \red{with respect to the laser's emission wavelength, as highlighted in Fig. 1(b).} \red{As a result, the feedback strength varies with the VOA voltage, but also with the detuning due to the FBG reflectivity profile. In addition, the VOA response has a sigmoid shape as shown in the supplementary information: transmission is maximal slightly above 4V and minimal slightly below 1V. We show the estimated feedback ratio for the whole feedback loop as a function of detuning for 4 different VOA voltages in Fig. 1(b).} In order to have a relatively high emitted power level, the injection current is set at $26$\,mA ($\sim 2.5 \times$the threshold current). The laser spectrum without feedback is shown in the inset of Fig. 1(c). The laser dynamics are characterized by the bandwidth at -$20$\,dB of the main peak evaluated based on optical spectral data as depicted in Fig. 1(c) for a broadened spectrum. The choice of this figure of merit is based on three important criteria: straightforward, systematic, and simple compared to the commonly used merits extracted from the time series processing. 

We show in Fig. 2(a) the bandwidth at -$20$\,dB as function of VOA voltage and frequency detuning. Since the bandwidth is intrinsically linked to the laser dynamics, we can easily observe the evolution of the laser dynamics. The dark blue represents the regions where the laser is in a stable state. While the most complex chaotic behavior --- which typically has the broadest bandwidth --- is displayed in yellow. The color gradient in between represents the gradual broadening of the optical spectrum between the two extreme cases. There is a correlation between the minimum feedback strength required for instability and the FBG refectivity spectrum thus, the re-stabilization regions at detuning around -$30$\,GHz results from the dip of the FBG reflectivity profile. In the same way the larger bandwidths at small detunings can be attributed to the higher reflectivity of the FBG's main lobe in comparison with that of the first side lobe. Though the evolution of the dynamics along the vertical axis i.e. for varying feedback strength, is fully in line with standard expectations, its evolution along the horizontal axis brings some surprises. Besides the improved stability observed around a detuning of -$30$\,GHz, a roughly periodic stability enhancement visible across the whole range of feedback strength is observed. To effectively link this feature to fluctuations of the feedback phase, we analyze the evolution of the optical spectrum along a horizontal line at low feedback rates as shown in Fig. 2(b). The corresponding VOA voltage at which the optical spectra are taken is $4.5$\,V, at a position where the laser exhibits stable behavior and the optical spectrum is not broadened in order to accurately track the position of the peaks (the red thread in Fig. 2(a)). We observe the emergence of a periodic wavelength shift, consistent with a feedback phase variation, well aligned with the recorded periodicity of the dynamical evolution. Furthermore, the wavelength of the laser (without feedback) remains untouched in the whole experiment as it is temperature controlled and pumped with fixed current. Hence, the wavelength variation can only be induced by the feedback phase changes. 
\begin{figure}[htbp]
\centering
\includegraphics[width=0.9\linewidth]{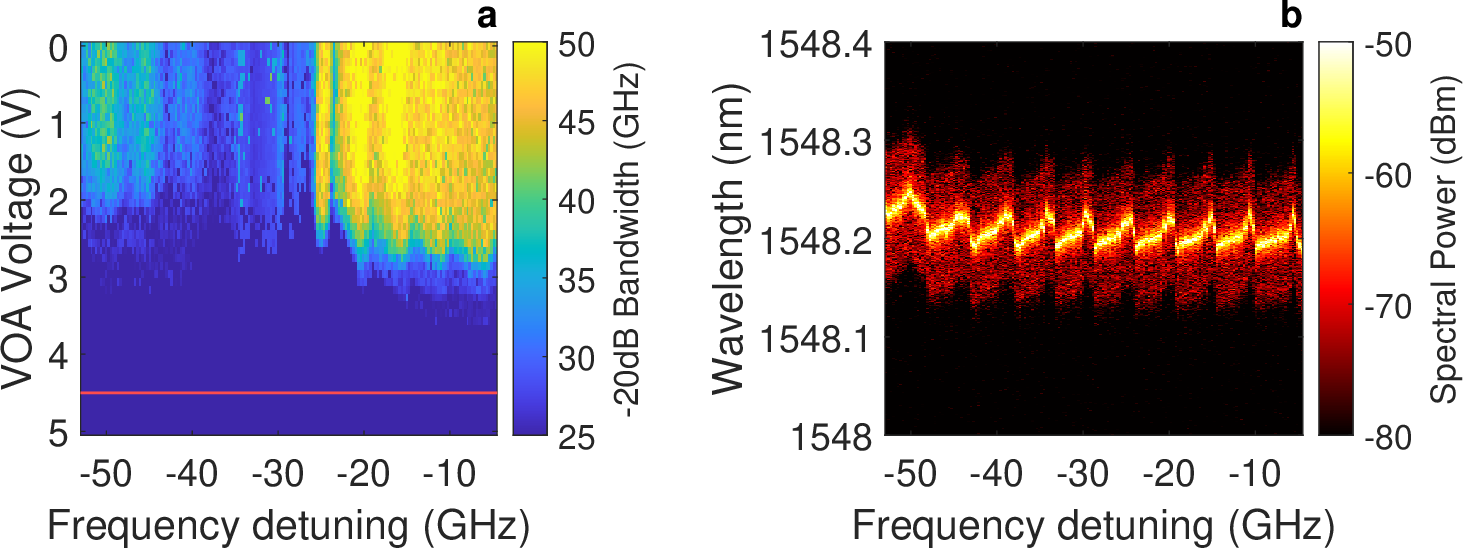}
\caption{(a) Experimental map of the values of bandwidth at -$20$\,dB as a function of frequency detuning and VOA voltage for a laser with feedback from a $4$\,mm long FBG. Note that the color scale is saturated at the lower end in order to make the changes more visible. (b) Optical spectra as function of frequency detuning at VOA voltage $4.5$\,V.}
\label{fig:Experiment_1}
\end{figure}

It has been shown that in optical fibers the optical phase fluctuates due to the thermal effects with a phase sensitivity of about 48 \,rad/m/K at 1550\,nm \cite{slavik_ultralow_2015}. Hence, any temperature change --- including localized changes --- will therefore lead to a variation of the feedback phase. As a result, in our experimental setup, temperature tuning of the FBG center wavelength will simultaneously lead to phase variations. We suppose that the total phase that the light accumulates while traveling in the feedback cavity is given by:
\begin{equation}
\theta = \theta_{TS}\Delta\Omega+\theta_0
\label{eq:phase_equation}
\end{equation}
where $\theta_{TS}$ is the temperature sensitivity phase and $\theta_0$ is the offset phase, that is, the initial phase of the system when the grating is not subjected to temperature variations. The FBG temperature tuning in our experiment produces  a change of $1.385$\,GHz/K of the FBG center frequency. Considering that the heated fiber portion is $3$\,cm it equates a temperature sensitivity phase change of $\approx 0.33 \pi$\,rad/GHz. 

Depending on the temperature gradient over the heated section of the fiber the periodic stability enhancement variations period is expected to change. To test that, the grating was positioned at three different positions along the heater, exposing different lengths of the fiber before and after the grating to a heat gradient. The heat-subjected fiber length changes proportionally from one position to the other \red{as illustrated in the inset of Fig. 1(a)}. After the grating position in the heater is adjusted, the wavelength detuning is varied via the temperature controller, as in the first experiment and all other parameters are the same. 

\begin{figure}[htbp]
     \centering
     \begin{subfigure}[b]{0.5\textwidth}
     \centering
        \includegraphics[width=0.8\linewidth]{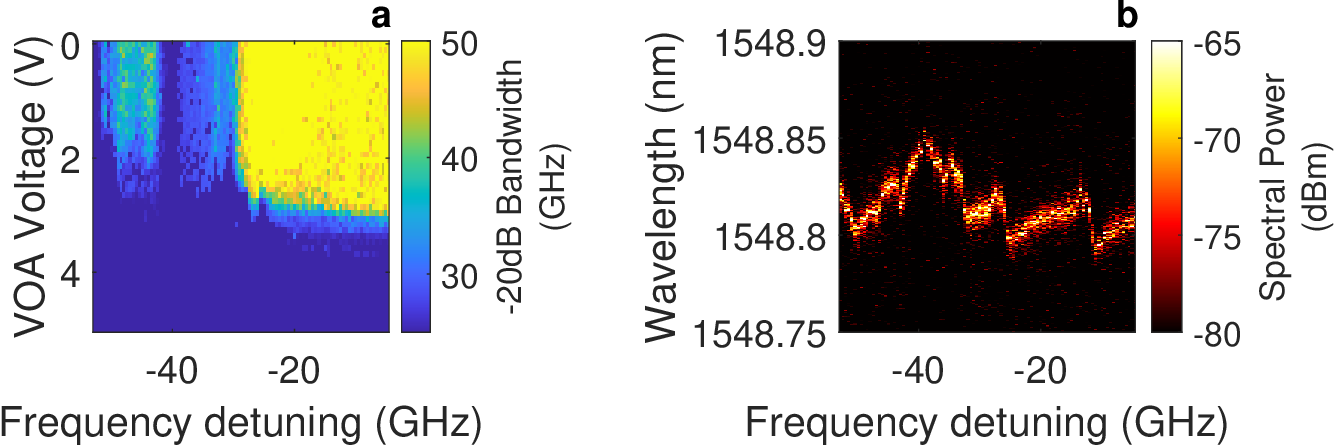}
     \end{subfigure}
     \hfill
     \begin{subfigure}[b]{0.5\textwidth}
     \centering
         \includegraphics[width=0.8\linewidth]{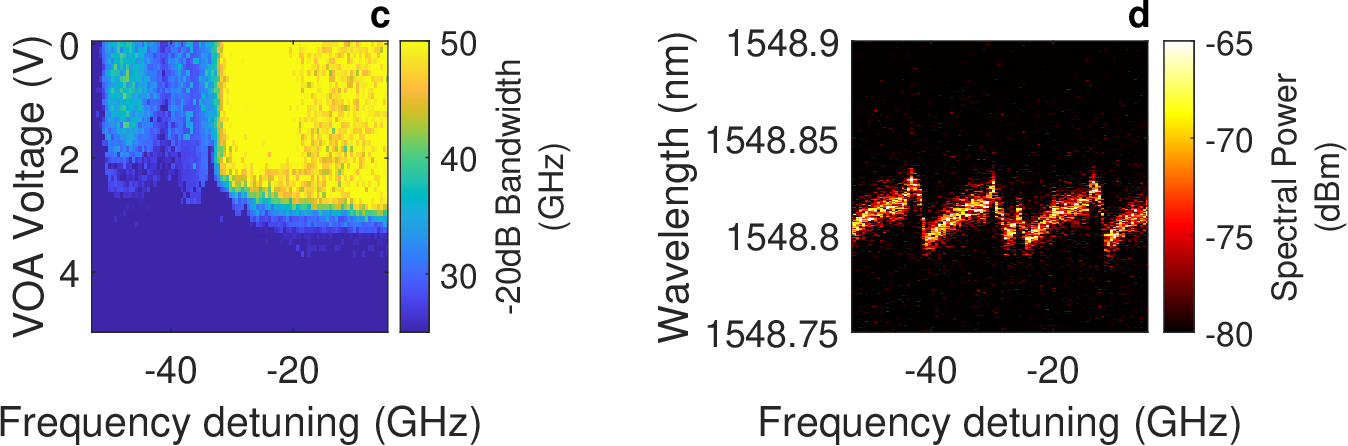}
     \end{subfigure}
     \begin{subfigure}[b]{0.5\textwidth}
     \centering
        \includegraphics[width=0.8\linewidth]{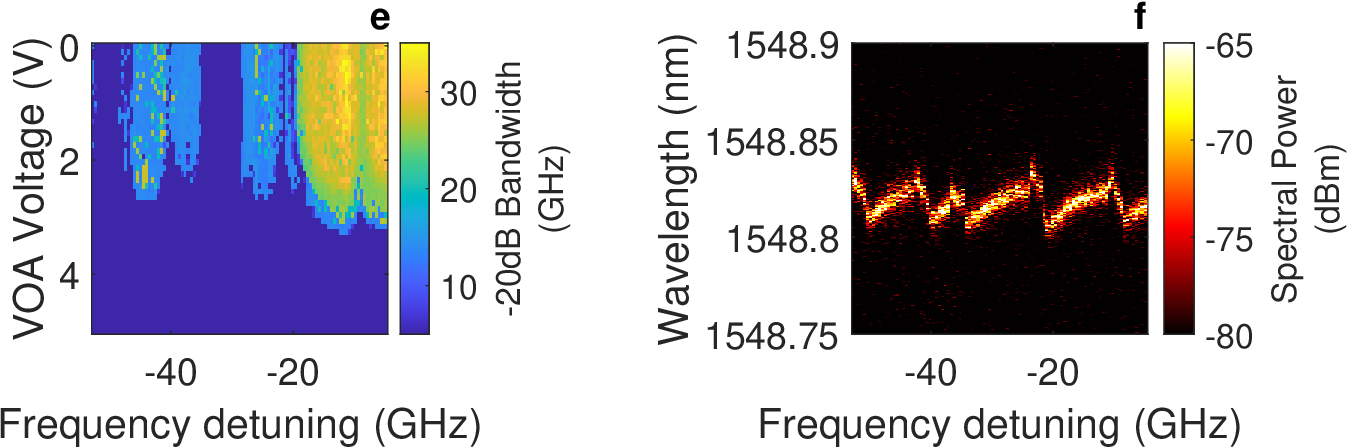}
     \end{subfigure}
        \caption{(a),(c),(e) Experimental map of the values of the bandwidth at -$20$\,\red{dB} as a function of frequency detuning and VOA Voltage for different positions of the grating on the heater. The optical spectra bandwidth for the third position is reduced thus, the the color scale in (e) is different. (b),(d),(f) The corresponding optical spectra as function of wavelength detuning at VOA voltage $4.5$\,V.}
        \label{fig:Experiment_2}
\end{figure}

The experimental maps of bandwidth at -$20$\,dB for the three positions are shown in Fig. 3(a), (c), (e) and the qualitative changes resulting from the phase difference can be distinguished. The first two positions seem to trigger roughly the same laser behavior. The destabilization starts at the same feedback strength and a correlation with the FBG reflection profile can be observed. Instead, in the third position, the feedback strength necessary to destabilize the laser is higher compared to the first two positions and the optical bandwidth for the same conditions is reduced (for this reason the color scale in Fig. 3(e) is different from Fig. 3(a), (c)). Although it seems intuitively hard to attribute such significant change of the overall dynamics only to a change of the phase, we obtain consistent measurements when reproducing the measurement across different days. The seemingly periodic stability enhancement due to the temperature sensitive phase, discussed previously, is not visible for the first two positions but more prominent for the third one where it seems that the phase variations have a greater impact. However, when taking into consideration the evolution of the emission wavelength with the frequency detuning, plotted in Fig. 3(b), (d), (f), these alternations are still visible and repeated with a slightly different period for the three positions. Hence, we associate the qualitative changes of the experimental maps to the effects arising from the offset phase variations, the second term in eq. \ref{eq:phase_equation}, and the behavior we observe in these maps is a combined effect of the offset phase and the temperature sensitive phase. Depending on the starting value of the offset phase the stability enhancement variations could be more prominent as in the third position or vanish as happens for position one and two. In the first position, for detuning between -$30$\,GHz and -$50$\,GHz the periodicity degrades. This could be attributed to the environmental disturbances such as temperature variations and vibrations, possibly bringing some instability to the frequency detuning and thus compromising the results. This is a factor that should be taken into consideration for all the experimental results.

To confirm our interpretation on the impact of the feedback phase, we perform numerical simulations based on a normalized version of the Lang-Kobayashi rate equations, modified to include the effect of time-distributed feedback from the FBG as follows \cite{PhysRevA.53.4429,6384650}:
\begin{align}
\frac{dE}{dt} &= (1+i\alpha)N(t)E(t)+\kappa e^{-i \theta\tau}\left[r(t)e^{-i\Delta\Omega t}\right]\ast E(t-\tau)
\label{eq:rate_equations1} \\
\frac{dN}{dt} &= \frac{1}{T}\left(P-N(t)-(1+2N(t))|E(t)|^2\right)
\label{eq:rate_equations2} \\
r(\Omega) &=\Omega_{BW}\left(2\Omega +i\sqrt{\Omega_{BW}^2-4\Omega^2}\coth{\frac{\pi\sqrt{\Omega_{BW}^2-4\Omega^2}}{2\Omega_l}}\right)^{-1}
\label{eq:impulse_response}
\end{align}
where $E(t)$ represents the normalized complex electrical field and $N(t)$ the normalized carrier density in the laser cavity. The equations are normalized in time by the photon lifetime and thus all time-related parameters are unitless. $T$ is the normalized carrier lifetime, $P$ is the dimensionless pump parameter, $P=0$ corresponds to the threshold, $\kappa$ is the feedback rate and $\alpha$ the linewidth enhancement factor. In the simulations the values corresponding to $T, P$ and $\alpha$ are $1000, 1$ and $3$ respectively. The feedback optical phase is denoted with $\theta$ and is given by eq. \ref{eq:phase_equation}. The FBG feedback is described by the last term in the field equation, expressing the impulse response of the FBG reflection by $r(t)$, the field amplitude coupling back to the laser is proportional to the convolution $r(t)\ast E(t-\tau)$. $r(t)$ for the FBG case is obtained from the inverse Fourier transform of the FBG reflection frequency response $r(\Omega)$ as reported in \cite{1233724,6320711,618322}. The FBG parameters are $\Omega_{BW}$ and $\Omega_l$ representing the FWHM bandwidth of the main lobe in the reflectivity profile and frequency gap between neighboring side lobes, respectively. The values of $\Omega_{BW}$ and $\Omega_l$ for the $4$\,mm long grating are $29.95$\,GHz and $25.86$\,GHz accordingly considering a photon lifetime of $1$\,ps. In the simulations the normalized feedback rate is switched from $0$ to $0.005$. The simulated FBG reflectivity spectrum is shown in black in Fig. 4(a) (together with the experimental spectrum in blue for comparison). The scanned range of FBG is $55$\,GHz covering the the main and the first side lobe of the FBG as highlighted in Fig. 4(a). The temperature sensitive phase changes by $\theta_{TS}=0.36\pi$\,rad/GHz and offset phase is taken $\theta_0=0$.  

\begin{figure}[htbp]
\centering
\includegraphics[width=0.85\linewidth]{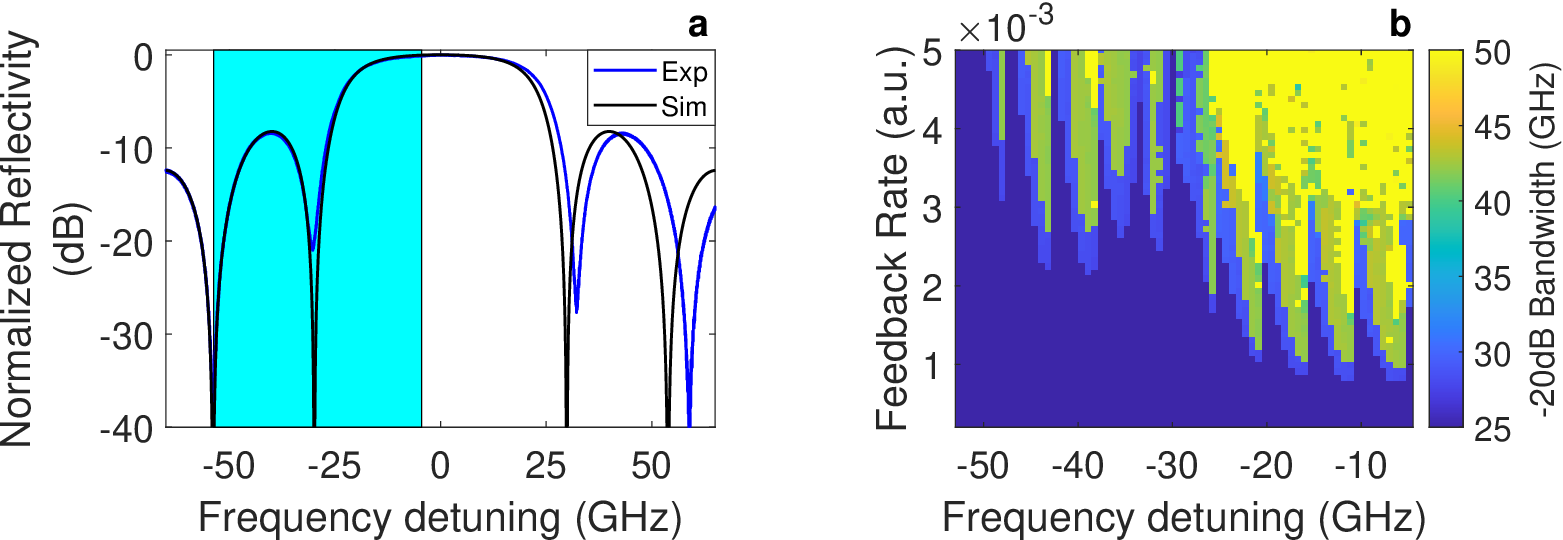}
\caption{(a) In black: the simulated normalized reflectivity spectrum of the FBG used throughout the simulation with corresponding  $\Omega_{BW}=29.95$\,GHz and $\Omega_l=25.86$\,GHz when considering the photon lifetime equal to $1$\,\red{ps}. In blue: the normalized reflectivity spectrum of the FBG used in the experiment. (b) Mapping of the bandwidth at -$20$\,\red{dB} in the parameter space of frequency detuning and feedback rate when $\theta_{TS}$ varies by $0.36\pi$\,rad/GHz and $\theta_0=0$.}
\label{fig:simulation_1}
\end{figure}

Fig. 4(b) shows the numerical mapping of the -$20$\,dB bandwidth in the frequency detuning and feedback rate plane. \red{For completeness, the evolution of the optical spectra for a low feedback rate can be found in the supplementary information.} We observe that the border between different dynamical regions qualitatively matches the behavior of the experimental maps. This consistent input indicates that the temperature sensitive phase is the underlying mechanism explaining the periodic stability enhancement. It is worth mentioning that here we focus in a small inset of the theoretical dynamical map reported in \cite{7097638}, where only the reflection phase of the FBG is considered. The FBG reflection phase exhibits rapid changes near the edges of the lobes corresponding to the chromatic dispersion as reported in \cite{618322} for different gratings. There the dynamics have been identified using the distribution of the time-series extrema however, the laser follows roughly the same behavior as the one we compute in Fig. 4(b) with the exception that the stability enhancement variations appear at the border between different states resulting from the extra phase term added on top of the FBG's reflection phase.

If we do the same fitting effort for the measurements shown in Fig. 3, that is considering also the effect of the phase offset $\theta_0$, then we obtain the results shown in Fig. 5. The offset phase $\theta_0$ is $\pi/4$\,rad, $\pi/2$\,rad and $\pi$\,rad at (a), (b) and (c) while the temperature sensitive phase $\theta_{TS}$ is equal to $0.027\pi$\,rad/GHz, $0.027\pi$\,rad/GHz and $0.182\pi$\,rad/GHz respectively. \red{Again, the corresponding evolution of the optical spectra for a low feedback rate can be found in the supplementary information.} As in the experimental results, the first two maps show roughly the same behavior, the border between the different dynamical states is smoother and the effect of the side lobe is visible where it seems that another threshold of the feedback rate is necessary to destabilize the laser. The effect of the temperature sensitive phase is only visible for detunings around -$45$\,GHz where a stability enhancement is visible around the middle of the side lobe. For the third map we observe the same bandwidth reduction resulting from the phase change consistent with the experimental data  (similar to Fig. 3(e) a different color scale is also used here). For this combination of offset and temperature sensitive phase the periodic stability enhancements are more prominent but with another period compared to Fig. 4(a). Overall, the simulated maps for consistent $\theta_{TS}$ and $\theta_0$ qualitatively match the experimental results illustrating the impact of the offset phase of the system and its combined effect with the temperature sensitive phase as given by Eq. \ref{eq:phase_equation}.

\begin{figure}[htbp]
\centering
\includegraphics[width=\linewidth]{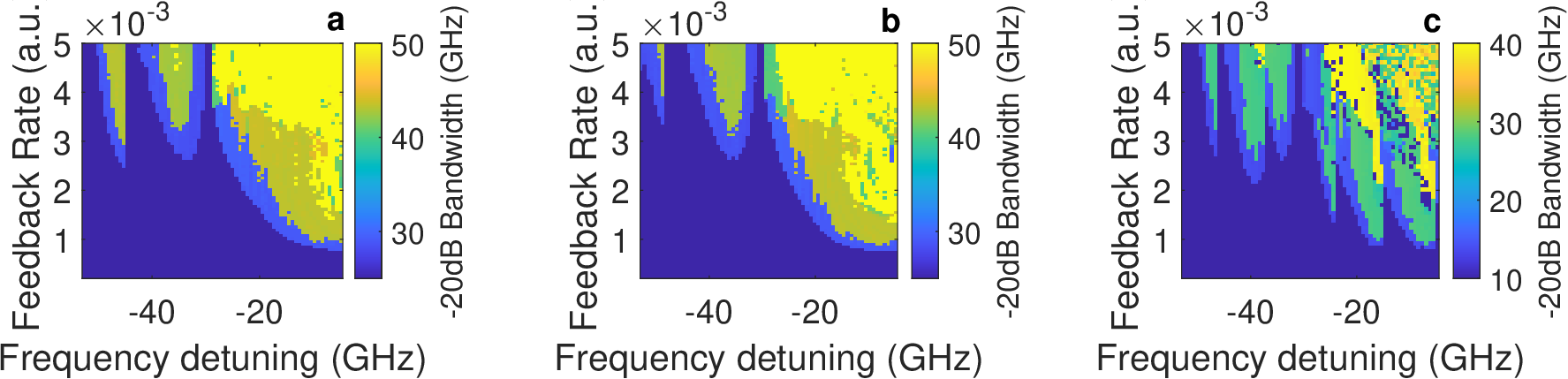}
\caption{Mapping of the bandwidth at -$20$\,\red{dB} in the parameter space of frequency detuning and feedback rate for (a) $\theta_{TS}=0.027\pi$\,rad/GHz; $\theta_0=\pi/4$\,rad (b) $\theta_{TS}=0.027\pi$\,rad/GHz; $\theta_0=\pi/2$\,rad  (c) $\theta_{TS}=0.182\pi$\,rad/GHz; $\theta_0=\pi$\,rad. The color scale for this map is different since the bandwidth values are lower.}
\label{fig:simulation_2}
\end{figure}

In conclusion, we have shown experimentally and numerically that semiconductor lasers have a strong sensitivity to the phase of time distributed feedback from an FBG. We considered both the effect of the temperature sensitive phase and the offset phase or the initial phase of the system. The temperature sensitive phase results from the thermal tuning of the FBG and triggers roughly periodic stability enhancement in the border between different dynamical states corresponding with periodic shifts of the emission wavelength consistent with the feedback phase variations. The offset phase of the system changes the overall dynamics exhibited by the laser influencing this way also the periodic stability enhancement variations. Depending on the initial conditions the variations could fade or become more prominent. The simulations are in good agreement with the experimental results, confirming once again the relevance of the Lang-Kobayashi model. Because such phase variation is likely to occur for any practical method employed to tune the Bragg Wavelength, such as temperature, strain or bending, we believe that this feature should be closely considered and monitored when dealing with FBG feedback.

\begin{backmatter}
\bmsection{Funding} Research Foundation - Flander (FWO, Fonds Wetenschappelijk Onderzoek). Projects: Post-Doc Scholarship of Martin Virte, REFLEX, project number 1530318N, FIONA, project number G029619N, COLOR'UP, project number G0G0319N

\bmsection{Disclosures} The authors declare no conflicts of interest.
\red{\bmsection{Supplemental document} See Supplement 1.} 
\end{backmatter}

\bibliography{sample}
\bibliographyfullrefs{sample}

\end{document}